\documentclass[aps, prd, onecolumn, tightenlines, notitlepage, superscriptaddress, nofootinbib, preprintnumbers, floatfix,showkeys]{revtex4-1}
\usepackage{amstext}
\usepackage{amssymb}
\usepackage{amsmath}
\usepackage{graphicx}
\usepackage{url}
\usepackage{color}
\usepackage{ulem}
\usepackage[utf8]{inputenc}
\pdfoutput=1
\usepackage{textcomp}

\usepackage{epsfig,amsfonts,mathrsfs,amsmath,amssymb,graphicx,color,slashed,multirow}
\usepackage{amsmath,latexsym,amssymb,graphicx,slashed,color,enumerate,url,cancel,gensymb}
\usepackage{textcomp}

\usepackage[x11names]{xcolor}
\usepackage[colorlinks]{hyperref}

\definecolor{nicered}{rgb}{0.7,0.1,0.1}
\definecolor{nicegreen}{rgb}{0.1,0.5,0.1}

\AtBeginDocument{\hypersetup{citecolor=Green3,linkcolor=Red1,urlcolor=blue}}

\begin{document}

\begin{flushright}
IFIC/20-19
\end{flushright}

\title{{\Large CPT and CP, an entangled couple}}

\author{Gabriela Barenboim}\email{gabriela.barenboim@uv.es}
\affiliation{Instituto de Física Corpuscular, CSIC-Universitat de València, C/Catedrático José Beltrán 2, Paterna 46980, Spain}
\affiliation{Departament de Física Teòrica, Universitat de València,
C/ Dr. Moliner 50, Burjassot 46100, Spain}
\author{Christoph A. Ternes}\email{chternes@ific.uv.es}
\affiliation{Instituto de Física Corpuscular, CSIC-Universitat de València, C/Catedrático José Beltrán 2, Paterna 46980, Spain}
\author{Mariam Tórtola}\email{mariam@ific.uv.es}
\affiliation{Instituto de Física Corpuscular, CSIC-Universitat de València, C/Catedrático José Beltrán 2, Paterna 46980, Spain}
\affiliation{Departament de Física Teòrica, Universitat de València,
C/ Dr. Moliner 50, Burjassot 46100, Spain}

\begin{abstract}
Even though it is undoubtedly very appealing to interpret the latest T2K results as evidence of CP violation, this claim assumes CPT conservation in the neutrino sector to an extent that has not been tested yet.
As we will show, T2K results are not robust against a CPT-violating explanation.
On the contrary, a CPT-violating CP-conserving scenario is in perfect agreement with current neutrino oscillation data.
Therefore, to elucidate whether T2K results imply CP or CPT violation is of utter importance.
We show that, even after combining with data from NO$\nu$A and from reactor experiments, no claims about CP violation can be made. Finally, we update the bounds on CPT violation in the neutrino sector.
\end{abstract}

%\keywords{a,b,c,d,.... }

\maketitle
%\newpage
%\tableofcontents

%%%%%%%%%%%%%%%%%%%%%%%%%%%%%%%%%
%%%%%%%%%%%%%%%%%%%%%%%%%%%%%%%%%
%%%%%%%%%%%%%%%%%%%%%%%%%%%%%%%%%
\section{Introduction}
%%%%%%%%%%%%%%%%%%%%%%%%%%%%%%%%%
%%%%%%%%%%%%%%%%%%%%%%%%%%%%%%%%%
%%%%%%%%%%%%%%%%%%%%%%%%%%%%%%%%%

The first hints for CP violation in the lepton sector appeared in global fits to neutrino oscillation data~\cite{deSalas:2017kay,Capozzi:2018ubv,Esteban:2018azc}. These hints were obtained only after combining the results from all experiments. None of them was disfavoring CP conservation on their own.
Recently, however, the T2K collaboration reported the observation of CP violation at approximately 3$\sigma$~\cite{Abe:2019vii}. 
Currently, NO$\nu$A's sensitivity~\cite{Acero:2019ksn} to measure the CP phase is not competitive with the one of T2K, although this should change in the upcoming years~\cite{Ghosh:2014dba}.
The T2K measurement is robust against several new physics scenarios. For example, in Refs.~\cite{Esteban:2019lfo} and~\cite{Miranda:2019ynh} it was shown that the sensitivity is not significantly reduced in the presence of neutrino non-standard interactions or for non-unitary neutrino mixing, respectively. 
Here, we will show that it is, however, not robust against CPT violation. The result, and therefore the claim that CP violation was found at any level, assumes that neutrino and antineutrino oscillation parameters in vacuum are identical (CPT conservation). 
This assumption is not supported by any experimental test and relies only in the "reasonable" expectation that Nature can be described in terms of local relativistic quantum field theory where CPT is built in. 
This may or may not be the case and, hence, the only way to make a solid statement about the violation of CP in the neutrino system at any rate must rule out the CPT violation possibility at the same level first.
Unfortunately, as we will show, this window is far from being closed in the light of current data.
If we assume neutrinos and antineutrinos oscillate with different parameters, we can obtain an equally good fit setting the CP phases for neutrinos and antineutrinos to any arbitrary value, but allowing for different mixing angles and mass splittings.
Therefore, the  claims in favor of the observation of CP violation would not stand anymore in this scenario. 
Although it will not be discussed in detail along this paper, this argument also applies to the recent hints in favor of normal mass ordering, obtained assuming CPT conservation, as well. 
Indeed, one should be very careful doing this assumption, since imposter solutions can be obtained, as discussed in Ref.~\cite{Barenboim:2017ewj}.
In order to confront our hypothesis not only with T2K, but also with all relevant neutrino oscillation data that play a role in the determination of CP violation, in this work we analyze the most recent long-baseline neutrino and antineutrino  data samples from T2K~\cite{Abe:2019vii} and NO$\nu$A~\cite{Acero:2019ksn} and the latest antineutrino data from the reactor experiments Daya Bay~\cite{Adey:2018zwh} and RENO~\cite{Bak:2018ydk}.

In Sec.~\ref{sec:t+n} we discuss the analysis of T2K and NO$\nu$A neutrino and antineutrino data under the assumption of CPT violation, this is, allowing for different oscillation parameters in each channel. 
Next, in Sec.~\ref{sec:lbl+reac} we perform a combined analysis of long-baseline and reactor data and discuss the implications of our results in connection with the recent T2K claim on the measurement of CP violation. 
The analysis performed in this paper gives us the chance to update the current bounds on CPT violation in the neutrino sector, that are presented in Sec.~\ref{sec:bounds}. Finally, we summarize and present our conclusions in Sec.~\ref{sec:conc}.

%%%%%%%%%%%%%%%%%%%%%%%%%%%%%%%%%
%%%%%%%%%%%%%%%%%%%%%%%%%%%%%%%%%
%%%%%%%%%%%%%%%%%%%%%%%%%%%%%%%%%
\section{Analysis of T2K and NO$\nu$A data}
\label{sec:t+n}
%%%%%%%%%%%%%%%%%%%%%%%%%%%%%%%%%
%%%%%%%%%%%%%%%%%%%%%%%%%%%%%%%%%
%%%%%%%%%%%%%%%%%%%%%%%%%%%%%%%%%

T2K collected a large data sample, corresponding to an exposure at Super-Kamiokande of 1.49$\times10^{21}$ protons on target (POT) in neutrino mode and 1.63$\times10^{21}$ POT in antineutrino mode~\cite{Abe:2019ffx,Abe:2019vii} observing 243 (140) muon (anti-muon) events and 75 (15) electron (positron) events. 
In addition, there are 15 electron events where also a pion is produced. 
These results improve the former release~\cite{Abe:2018wpn}, allowing now to exclude CP-conserving values of $\delta$ at close to 3$\sigma$ confidence level, when assuming CPT invariance. 
The same parameters as in T2K are measured by the NO$\nu$A experiment. NO$\nu$A collected data corresponding to 8.85$\times10^{20}$ POT in neutrino mode~\cite{NOvA:2018gge} and 12.33$\times10^{20}$ POT in antineutrino mode~\cite{Acero:2019ksn}. 
NO$\nu$A observes 113 (102) muon (anti-muon) events in the disappearance channel, expecting 730 (476) without oscillations, and 58 (27) electron (positron) events in the appearance channel. 
The 27 events in the antineutrino mode consist the first ever significant observation of $\overline{\nu}_e$ appearance in a long-baseline experiment~\cite{Acero:2019ksn}. 
Unlike in the case of T2K, the NO$\nu$A best fit value is obtained for $\delta = 0$, resulting in a small tension with the T2K result. 
However, the T2K measurement of the CP phase is statistically more relevant than the corresponding NO$\nu$A result.

\begin{figure}[t!]
\centering
\includegraphics[width=0.8\textwidth]{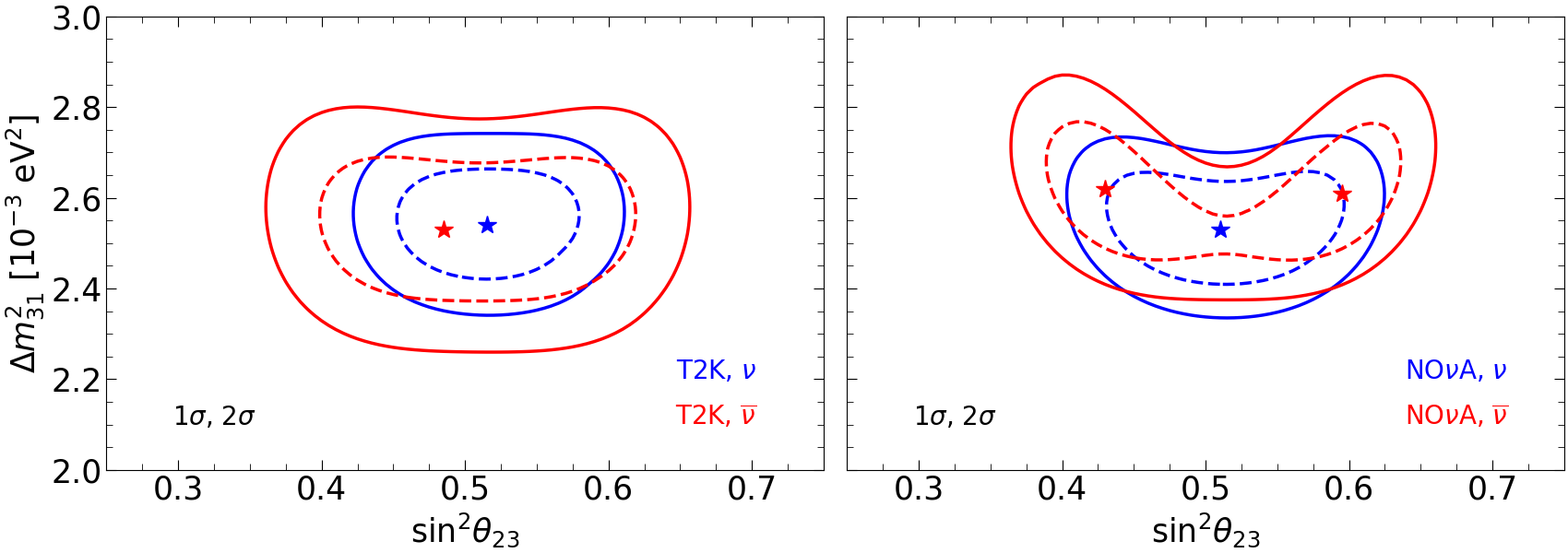}
\caption{1$\sigma$ (dashed) and 2$\sigma$ (solid)  allowed regions in the $\sin^2\theta_{23}$--$\Delta m_{31}^2$ plane 
($\sin^2\overline{\theta}_{23}$--$\Delta \overline{m}_{31}^2$ plane for antineutrinos)
for T2K (left) and NO$\nu$A (right) neutrino (blue) and antineutrino (red) data. The stars correspond to the best fit values obtained in each analysis.}
\label{fig:TN_sq23_dm31}
\end{figure}

\begin{figure}[t!]
\centering
\includegraphics[width=0.8\textwidth]{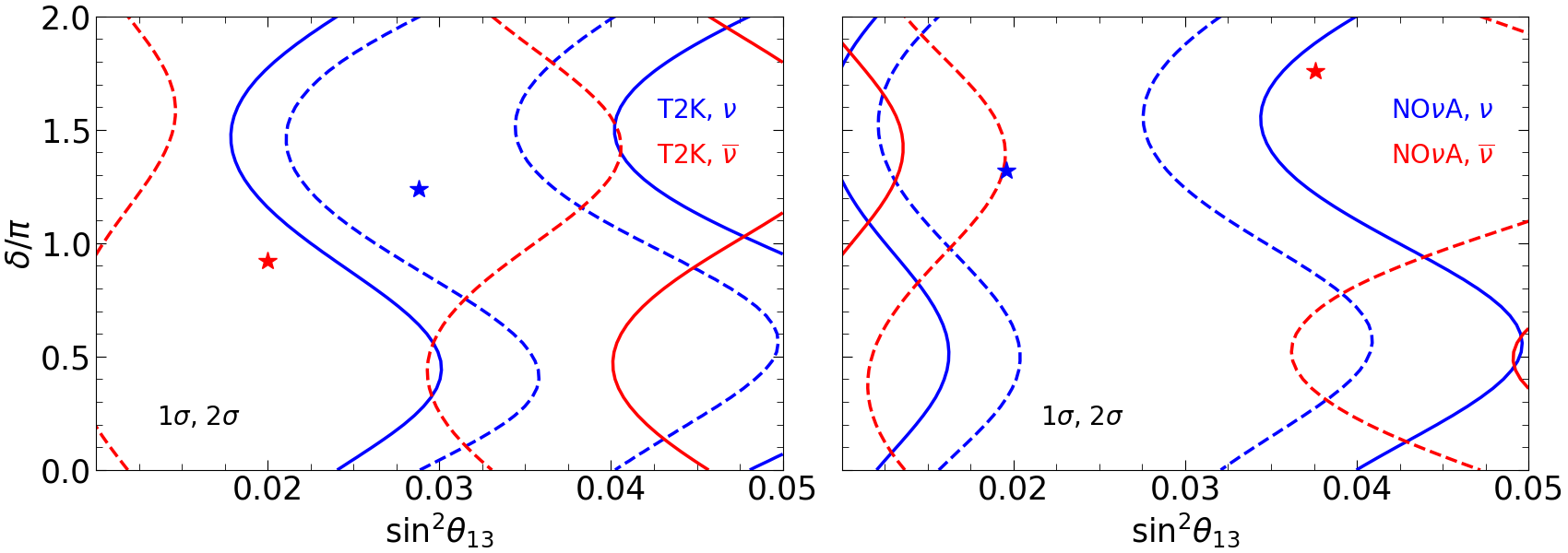}
\caption{1$\sigma$ (dashed) and 2$\sigma$ (solid) allowed regions in the $\sin^2\theta_{13}$--$\delta$ plane ($\sin^2\overline{\theta}_{13}$--$\overline{\delta}$ plane for antineutrinos) for T2K (left) and NO$\nu$A (right) neutrino (blue) and antineutrino (red) data. The stars correspond to the best fit values obtained in each analysis.}
\label{fig:TN_sq13_delta}
\end{figure}

\begin{figure}[t!]
\centering
\includegraphics[width=0.4\textwidth]{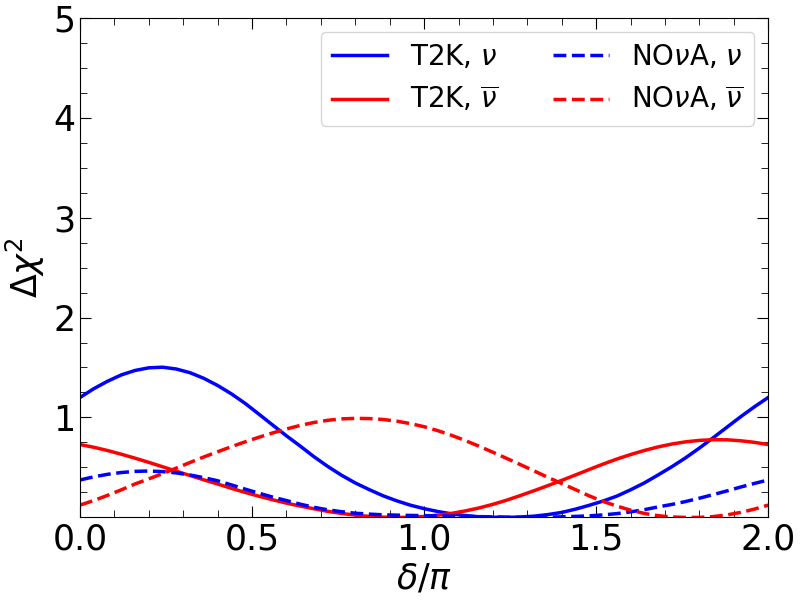}
\caption{$\Delta\chi^2$ profiles obtained from the analysis of neutrino (blue) and antineutrino (red) data from T2K (solid) and NO$\nu$A (dashed) for the CP phases $\delta$ (neutrinos) and $\overline{\delta}$ (antineutrinos).}
\label{fig:T_N_profiles}
\end{figure}

Here, we analyze neutrino and antineutrino data separately. 
In order to perform this analysis, we first make sure that our analysis reproduces well the published results in the CPT-conserving case. 
All necessary information on the experimental details is extracted from the corresponding references for each experiment, Refs.~\cite{Abe:2019ffx,Abe:2019vii,NOvA:2018gge,Acero:2019ksn} for the accelerators and Refs.~\cite{Adey:2018zwh,An:2016ses,An:2016srz,Bak:2018ydk,Seo:2016uom,Ahn:2010vy} for the reactors, to be discussed in the next section.
We extracted also the data points, expected event spectra, backgrounds and information on the systematic uncertainties from these references.
After reproducing the results obtained by the experimental collaborations for each experiment, we perform the analysis of neutrino and antineutrino data separately.
Since the effect of solar neutrino oscillation parameters is not appreciable in the experiments discussed in this paper, we keep them fixed throughout our analysis at $\sin^2\theta_{12} = \sin^2\overline{\theta}_{12} = 0.32$ and $\Delta m_{21}^2 = \Delta \overline{m}_{21}^2 = 7.55\times10^{-5}$~eV$^2$~\cite{deSalas:2017kay}. Here $x$ and $\overline{x}$ refer to neutrino and antineutrino oscillation parameters, respectively.
In addition, we assume normal neutrino and antineutrino mass ordering throughout the paper. Note that, although this possibility is not considered here,  different orderings for neutrinos and antineutrinos would already be an indication of CPT violation. The experiments discussed here have, however, only a limited sensitivity to measure the mass ordering on their own, which is further decreased when relaxing CPT conservation.
Our results are shown in Figs.~\ref{fig:TN_sq23_dm31}, \ref{fig:TN_sq13_delta} and \ref{fig:T_N_profiles}. 
Note that we always fit all  the parameters at the same time. In all the figures the parameters not plotted have been marginalized over.
We will start discussing the separate  neutrino and antineutrino analyses of T2K and NO$\nu$A in the atmospheric plane $\sin^2\theta_{23}$--$\Delta m^2_{31}$ ($\sin^2\overline{\theta}_{23}$--$\Delta\overline{m}^2_{31}$ for antineutrinos), shown in Fig.~\ref{fig:TN_sq23_dm31}.
In the left panel, we observe a very good agreement in the regions preferred by T2K neutrino and antineutrino data.
As expected, the sensitivity is much better in the neutrino channel, but still one can appreciate a total overlap between the two regions and very close values for the best fit points obtained in both analysis.
This improves a former result published by the T2K Collaboration, which observed somewhat different values for the atmospheric  mixing angles~\cite{Abe:2017bay}.
Note, however, that,  although the region for antineutrinos is now smaller than in  this former analysis, the old best fit values $\sin^2\overline{\theta}_{23} = 0.42$ and 0.59 still remain inside the 1$\sigma$ region.
Future oscillation data from DUNE could improve the measurement of $\theta_{23}$ and $\overline{\theta}_{23}$ considerably, being able  to single out the neutrino and antineutrino parameters in case they are truly different, as discussed in Refs.~\cite{Barenboim:2017ewj} and~\cite{Barenboim:2018lpo}.
Interestingly, the benchmark point we chose there, inspired  by the T2K measurement in Ref.~\cite{Abe:2017bay}, corresponds now with the best fit point obtained by NO$\nu$A and, therefore, remains a valid choice of benchmark point.
Moving to the  right panel of Fig.~\ref{fig:TN_sq23_dm31}, we find a different picture.
In particular,  we  see that NO$\nu$A shows  a disagreement in the best fit values of $\Delta m_{31}^2$ and $\sin^2\theta_{23}$ obtained for neutrinos and antineutrinos.
Actually, one observes that the values of $\sin^2\overline{\theta}_{23}$ further away from maximal mixing are correlated to larger values of $\Delta \overline{m}_{31}^2$.
Then, in the light of the new data,  one can conclude that  NO$\nu$A results will be more in favor of a CPT-violating description, which reinforces the legitimacy of our approach.
%
%% Fig 2 %%
%
Regarding the reactor angle, a small tension between the neutrino and antineutrino data analyses appears for T2K, as one can appreciate in the left panel of Fig.~\ref{fig:TN_sq13_delta}.
There, it is clearly visible that the wavy structure that defines the allowed region in the $\sin^2\theta_{13}$--$\delta$ plane is shifted towards larger values of $\theta_{13}$ for neutrinos, while the lines corresponding to the antineutrino mode are centered around $\sin^2\overline{\theta}_{13} \simeq 0.02$, the  best fit value for reactor neutrinos, as we will see in the next section.
In the case of NO$\nu$A, as illustrated in the right panel of Fig.~\ref{fig:TN_sq13_delta}, we find a slightly better agreement in the analysis of neutrino and antineutrino data, in part due to the poorer sensitivity to this mixing angle in comparison to T2K.
%
%%% Fig 3 %%
%
If we now turn to the sensitivity to the CP phases, $\delta$ and $\overline{\delta}$, we see  in Fig.~\ref{fig:TN_sq13_delta}, but specially in  the marginalized $\Delta\chi^2$ profiles shown in Fig.~\ref{fig:T_N_profiles}, that all values of the CP phases remain allowed at approximately 1$\sigma$ for both experiments.
This means that, unfortunately, at present, neither T2K or NO$\nu$A alone can make any significant statement about the measurement of the CP phase  without assuming CPT invariance. 
To understand the  main difference between our results and the ones obtained with the usual CPT-conserving analyses, one should recall the origin of the sensitivity to the CP phase in those analyses. 
If we assume CPT invariance, for a fixed value of $\theta_{13}$ (common to neutrinos and antineutrinos), the presence of a non-zero CP phase can induce a shift in the neutrino oscillation probability (and therefore in the event numbers) into different directions for neutrinos and antineutrinos. 
This behavior is illustrated in Fig.~1 of Ref.~\cite{Abe:2019vii}.
If we allow, however, for the angles to be different ($\theta_{13} \neq \overline{\theta}_{13}$), there is no need for invoking non-zero values of the CP-violating phase to reproduce the observed number of neutrino and antineutrino events, since the mixing angles can directly be adjusted to reproduce the experimental results.
As a result, the sensitivity to CP violation in the CPT-violating scenario is very poor and, therefore, much more statistics would be necessary to disentangle the effects of $\theta_{13}$ and $\delta$ using only the neutrino or the antineutrino channel.
It is then not surprising that we can not measure the CP phase in the separated analysis of T2K and NO$\nu$A data.

%%%%%%%%%%%%%%%%%%%%%%%%%%%%%%%%%
%%%%%%%%%%%%%%%%%%%%%%%%%%%%%%%%%
%%%%%%%%%%%%%%%%%%%%%%%%%%%%%%%%%
\section{Combined analysis of accelerator and reactor data}
\label{sec:lbl+reac}
%%%%%%%%%%%%%%%%%%%%%%%%%%%%%%%%%
%%%%%%%%%%%%%%%%%%%%%%%%%%%%%%%%%
%%%%%%%%%%%%%%%%%%%%%%%%%%%%%%%%%

The main goal of this section is to estimate the sensitivity to the CP-violating phases $\delta$ and $\overline{\delta}$ by performing combined analyses of all relevant neutrino and antineutrino data. 
As we have discussed before, one of the main limitations to the measurement of the CP phase in CPT-violating scenarios is the existence of  correlations between the phase $\delta$ ($\overline{\delta}$) and the mixing angle $\theta_{13}$ ($\overline{\theta}_{13}$) in the neutrino (antineutrino) channel, that can not be broken assuming identical parameters, as it happens in the CPT-conserving case.
These correlations can be disentangled (at least to some extent) by performing combined analyses with other complementary data samples, in analogy with the strategy behind  global fits to neutrino oscillation data~\cite{deSalas:2017kay,Capozzi:2018ubv,Esteban:2018azc}. 
Since, at present, long-baseline  experiments provide the world leading measurements for all the parameters under study  except for the reactor angle $\overline{\theta}_{13}$, here we will perform combined analyses of long-baseline and reactor oscillation data. 
Specifically, we will combine the most recent results from T2K and NO$\nu$A (already analyzed in Sec.~\ref{sec:t+n}.) with the latest data from the short-baseline reactor experiments Daya Bay~\cite{Adey:2018zwh} and RENO~\cite{Bak:2018ydk}.
In the case of neutrinos, this joint analysis corresponds just to the combination of T2K and NO$\nu$A data, while for the antineutrino channel we combine the results of the four experiments: T2K, NO$\nu$A, Daya Bay and RENO.
To highlight the impact of reactor data in the combined analysis,  we will also present and discuss the results corresponding to  accelerator data only in the case of antineutrinos.
\begin{figure}[t!]
\centering
\includegraphics[width=0.8\textwidth]{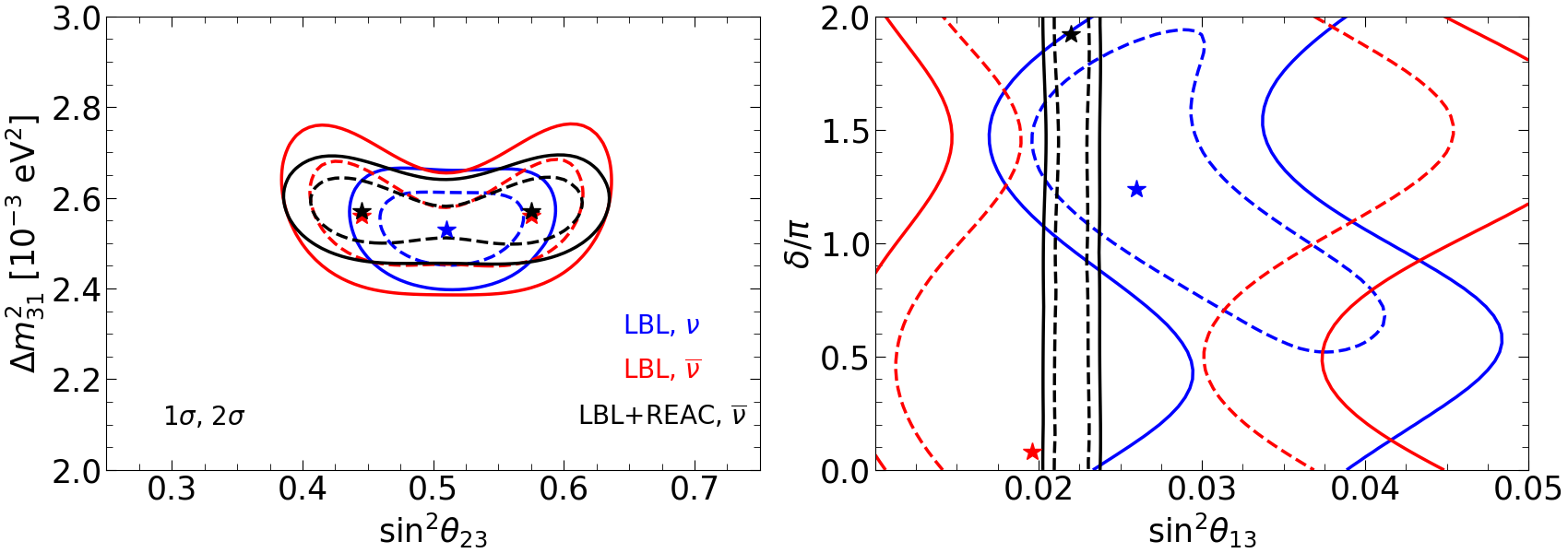}
\caption{1$\sigma$ (dashed) and 2$\sigma$ (solid) allowed two-dimensional projections from the  combined analysis of neutrino data (blue), antineutrino data from long-baseline accelerators alone (red) and combining with reactors (black). The stars correspond to the best fit values obtained in each analysis.}
\label{fig:comb_2D}
\end{figure}
\begin{figure}[t!]
\centering
\includegraphics[width=0.8\textwidth]{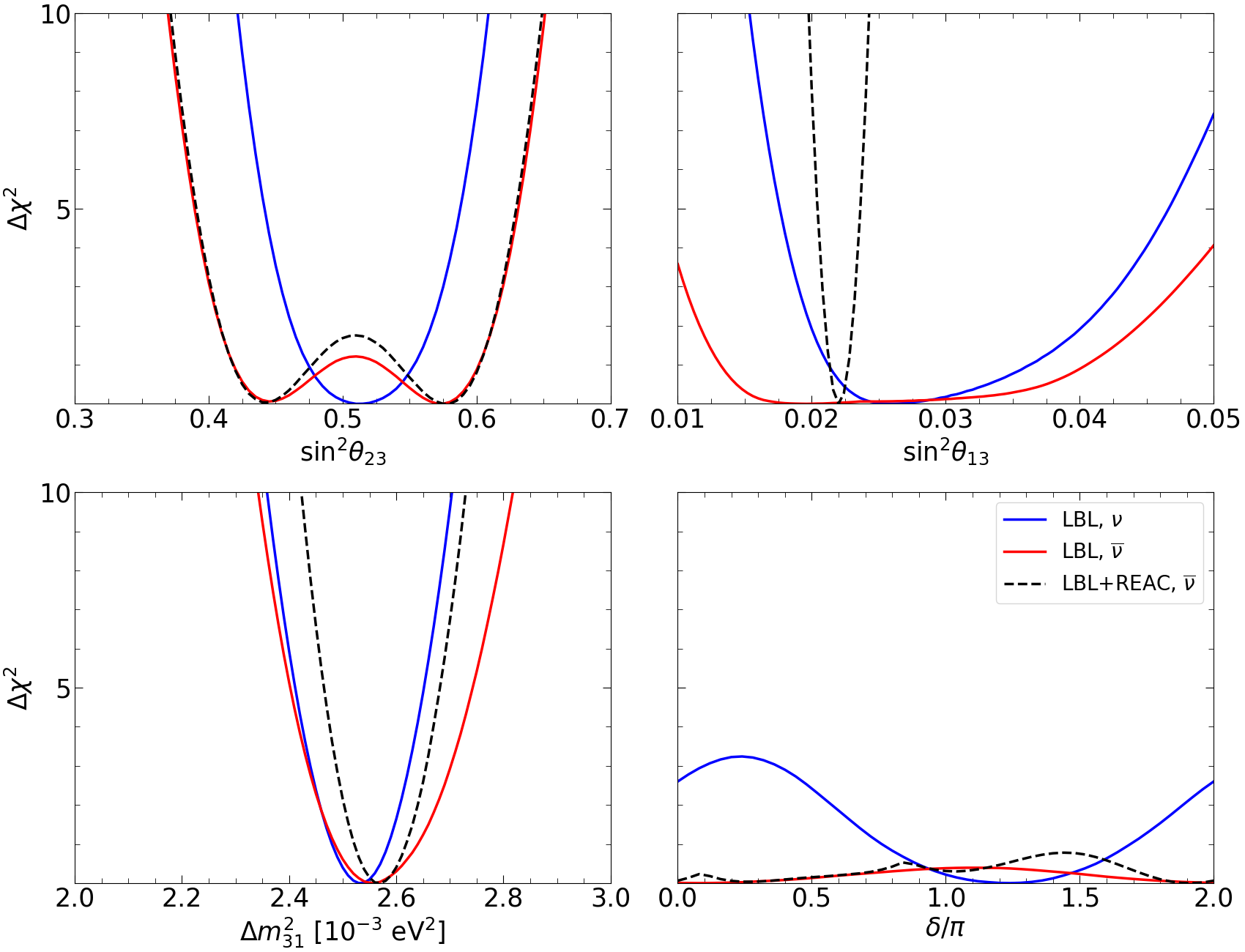}
\caption{$\Delta\chi^2$ profiles from the combined analysis of neutrino data (blue), antineutrino data from long-baseline accelerators alone (red) and combining with reactors (black dashed) for the atmospheric mixing angles (upper left), reactor mixing angles (upper right), atmospheric mass splittings (lower left) and CP phases (lower right).}
\label{fig:comb_profiles}
\end{figure}

Our results are shown in Figs.~\ref{fig:comb_2D} and \ref{fig:comb_profiles}.
The two-dimensional allowed regions for the oscillation parameters obtained from these analyses are shown in Fig.~\ref{fig:comb_2D}, while Fig.~\ref{fig:comb_profiles} shows the independent $\Delta\chi^2$ profiles corresponding to each parameter. As before, the undisplayed parameters have been marginalized over.
%
% atmos parameters
%
Regarding the determination of the atmospheric parameters (left panels of the figures), we find that, as one could expect from the  separate analyses of T2K and NO$\nu$A in Fig.~\ref{fig:TN_sq23_dm31}, the combination of both experiments prefers maximal mixing for the neutrino mixing angle $\theta_{23}$. In the antineutrino channel, on the contrary, the combined analysis of long-baseline data prefers the non-maximal values $\sin^2\overline{\theta}_{23}$ = 0.45 and 0.58, although maximal mixing is allowed with $\Delta\chi^2 = 1.2$.
After combining with reactors, the rejection against maximal mixing is further increased a bit. In any case, the result is not very significant, and maximal $\overline{\theta}_{23}$ remains allowed with $\Delta\chi^2 = 1.7$.
In the determination of the atmospheric mass splittings, we find that neutrino data prefer a lower best fit value, $\Delta m_{31}^2 = 2.53\times 10^{-3} ~\mathrm{eV}^2$, while the somewhat larger value of $\Delta\overline{m}^2_{31}$ favored by NO$\nu$A antineutrino data, shown in the right panel of Fig.~\ref{fig:TN_sq23_dm31}, shifts the preferred value in the combined long-baseline antineutrino analysis to $\Delta\overline{m}^2_{31} = 2.56\times 10^{-3} ~\mathrm{eV}^2$.
After combining with reactor experiments, the best fit value moves  to $\Delta\overline{m}^2_{31} = 2.57\times 10^{-3} ~\mathrm{eV}^2$. 
While the best fit point is only slightly shifted, the combination with reactor data results in an increased sensitivity to the mass splitting, better than the one obtained in the neutrino mode, due to the competitive measurement of $\Delta\overline{m}^2_{31}$ achieved at reactor experiments.
%
%%% theta_13 %%%
%
As it is widely known, the mixing angle $\overline{\theta}_{13}$ is best measured by reactor experiments.
This can be clearly seen in the right panel of Fig.~\ref{fig:comb_2D}, as well as in the upper right panel of Fig.~\ref{fig:comb_profiles}, where the contribution from long-baseline accelerators to this measurement is basically negligible. Unfortunately, the lack of a complementary measurement in the neutrino mode results in a much worse sensitivity in the determination of the neutrino angle $\theta_{13}$.
%
%%%  delta %%%%
%
Finally, the sensitivity to the CP phases from the combined analysis is presented in  the right panel of Fig.~\ref{fig:comb_2D} and the lower right panel of Fig.~\ref{fig:comb_profiles}. 
Compared to the separate analyses shown in Fig.~\ref{fig:T_N_profiles}, we see an improvement in the determination of $\delta$, together with a worsening in the sensitivity to $\overline{\delta}$. 
This is due to the fact that T2K and NO$\nu$A analyses show similar results in neutrino mode (see blue lines in Fig.~\ref{fig:T_N_profiles}), while for antineutrinos the region around the best fit value obtained by T2K is maximally penalized by the NO$\nu$A measurement and vice versa, as indicated by the red lines in Fig.~\ref{fig:TN_sq13_delta}.
This result barely improves after combining with reactors, as shown by the black dashed lines. 
From the combined T2K + NO$\nu$A neutrino analysis, we see that the CP-conserving value $\delta = 0$ is disfavored with $\Delta\chi^2 = 2.6$, while the other CP-conserving value,  $\delta = \pi$, is in much better agreement with data, with $\Delta\chi^2 = 0.2$. 
Regarding the antineutrino combined analysis, all possible values of the CP phase $\overline{\delta}$ remain allowed with $\Delta\chi^2<1$, even after combining with reactor data. 

In summary, we have observed that the combination of several data samples improves a bit the sensitivity to some of the oscillation parameters, but not at the level required to make claims, particularly in the case of the CP phases. As we have shown, the impact of reactor data is especially important in the determination of $\sin^2\overline{\theta}_{13}$ and $\Delta \overline{m}_{31}^2$, while it is not very significant in the determination of $\sin^2\overline{\theta}_{23}$ and $\overline{\delta}$, where it enters only via correlations.
Therefore, we can conclude that,  even the combination of T2K, NO$\nu$A and short-baseline reactor data, if analyzed without assuming CPT conservation, can not exclude any value of the CP-violating phase above the 2$\sigma$ level.

%%%%%%%%%%%%%%%%%%%%%%%%%%%%%%%%%
%%%%%%%%%%%%%%%%%%%%%%%%%%%%%%%%%
%%%%%%%%%%%%%%%%%%%%%%%%%%%%%%%%%
\section{Updated bounds on CPT violation}
\label{sec:bounds}
%%%%%%%%%%%%%%%%%%%%%%%%%%%%%%%%%
%%%%%%%%%%%%%%%%%%%%%%%%%%%%%%%%%
%%%%%%%%%%%%%%%%%%%%%%%%%%%%%%%%%

\begin{figure}[t!]
\centering
\includegraphics[width=1\textwidth]{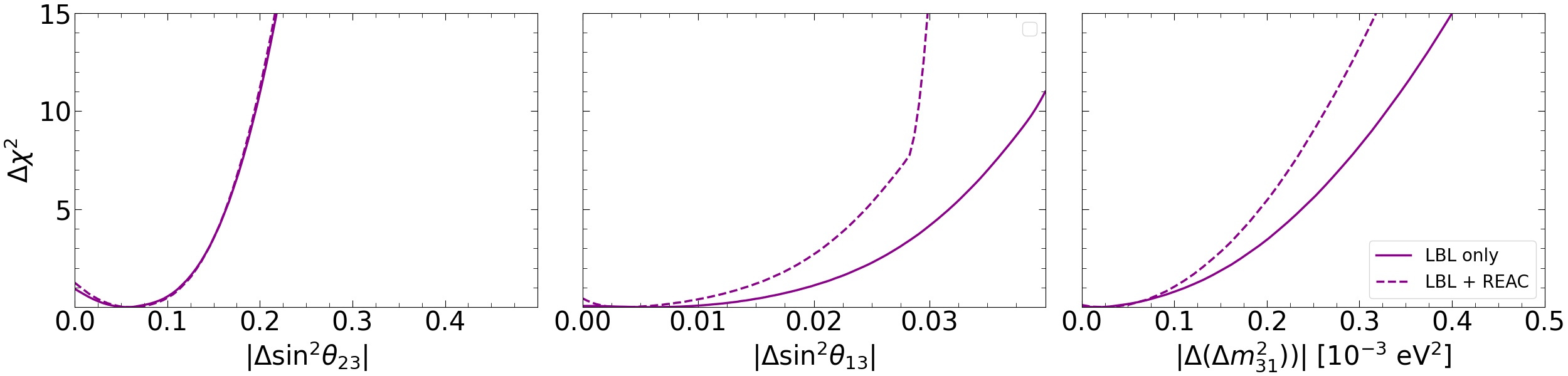}
\caption{$\Delta\chi^2$ profiles for the CPT-violating variables as obtained from the analysis of long-baseline accelerator data (solid) and from the combined analysis of accelerator and reactor experiments (dashed).}
\label{fig:Dx_profiles}
\end{figure}

We have discussed that, currently, a CPT-conserving and CP-violating measurement or a CP-conserving and CPT-violating measurement give both good fits to the data. 
However, this is not to be interpreted as an indication of CPT violation. 
Only a clear measurement of different mixing angles or masses for neutrinos and antineutrinos could be interpreted as a signal of CPT violation. 
Using the data discussed here, we can update the bounds we previously derived in Ref.~\cite{Barenboim:2017ewj}.
The new sensitivity on CPT violation is shown in Fig.~\ref{fig:Dx_profiles}, from where we can read off the current bounds on the differences $|\Delta x| = |x -\overline{x}|$. The most up-to-date bounds on CPT violation in the neutrino sector at 3$\sigma$ are
 \begin{eqnarray}
 & |\Delta m_{21}^2-\Delta \overline{m}_{21}^2| &< 4.7\times 10^{-5} ~\text{eV}^2,
  \nonumber \\
  & |\Delta m_{31}^2-\Delta \overline{m}_{31}^2| &< 2.5\times 10^{-4} ~\text{eV}^2,
 \nonumber \\
  & |\sin^2\theta_{12}-\sin^2\overline{\theta}_{12}| &< 0.14,
  \\
  & |\sin^2\theta_{13}-\sin^2\overline{\theta}_{13}| &< 0.029,
  \nonumber \\
  & |\sin^2\theta_{23}-\sin^2\overline{\theta}_{23}| &< 0.19\nonumber .
 \label{eq:new-bounds}
 \end{eqnarray} 
Note that, since there are no new data in the solar sector, here we adopted the bounds on the solar parameters from Ref.~\cite{Barenboim:2017ewj}.
Regarding the reactor mixing angle, we also find that the bound on $|\sin^2\theta_{13}-\sin^2\overline{\theta}_{13}|$ remains mostly unchanged despite having included the latest data from Daya Bay and RENO. This comes from the fact that reactor experiments can only contribute to the determination of the antineutrino mixing angle $\overline{\theta}_{13}$, while the measurement of its neutrino counterpart, $\theta_{13}$, has been almost unaffected by the new oscillation data.
Fortunately, thanks to the most recent long-baseline data analyzed in this work, some improvement has been obtained in the bounds of the atmospheric parameters.
In particular, the constraint on the difference in atmospheric mass splittings, $|\Delta m_{31}^2-\Delta \overline{m}_{31}^2|$,  is improved from $3.7\times 10^{-4} ~\text{eV}^2$ to $2.5\times 10^{-4} ~\text{eV}^2$, while the bound on $|\sin^2\theta_{23}-\sin^2\overline{\theta}_{23}|$ has been improved by a factor of 2, approximately. 
In this last case, however, it should be noted that the limit is somewhat ``unstable''. 
The worse bound derived in Ref.~\cite{Barenboim:2017ewj} was due to the fact that both neutrino and antineutrino data were preferring atmospheric mixing angles quite far away from maximal mixing, allowing for degenerate solutions in both octants.
On the contrary, the new neutrino data sample analyzed here is more in agreement with maximal mixing. Therefore, the $\Delta\chi^2$ profile for neutrinos does not have a second minimum, resulting in a stronger limit on the difference in the atmospheric mixing angles $|\Delta\sin^2\theta_{23}|$.
Note that these constraints can be further improved in the future by DUNE, as discussed in Refs.~\cite{Barenboim:2017ewj,deGouvea:2017yvn,Abi:2018dnh,Abi:2020evt}.

%

%%%%%%%%%%%%%%%%%%%%%%%%%%%%%%%%%
%%%%%%%%%%%%%%%%%%%%%%%%%%%%%%%%%
%%%%%%%%%%%%%%%%%%%%%%%%%%%%%%%%%
\section{Conclusions}
\label{sec:conc}
%%%%%%%%%%%%%%%%%%%%%%%%%%%%%%%%%
%%%%%%%%%%%%%%%%%%%%%%%%%%%%%%%%%
%%%%%%%%%%%%%%%%%%%%%%%%%%%%%%%%%

In this work we showed that, despite how tempting and exciting it could be to claim the experimental evidence of CP violation in the neutrino system, we are yet not close to do so. 
CPT violation (a more interesting phenomenon which challenges our description of Nature in terms of local relativistic quantum field theory) steps in the CP way.  
A difference between the oscillation parameters in the neutrino and antineutrino sector is fully consistent with all available neutrino data so far, and can account for the T2K and NO$\nu$A observations.
It is also important to stress that Occam's razor is of dubious application in this case, as CPT conservation is not an addition to the physics of the system but it is the cornerstone above which the whole system is built. 
Therefore, before any claim can be made about one particularly interesting feature of our phenomenological descriptions of neutrino oscillations, the tools used to construct it should be checked first.
As a step towards establishing the needed bounds on CPT (to claim a discovery of CP violation), we have updated the current limits on CPT invariance violation in the neutrino sector. 
The analysis of the new T2K and NO$\nu$A neutrino and antineutrino data samples in combination with reactor data has improved the limits on the differences of the atmospheric parameters with respect to our previous results in Ref.~\cite{Barenboim:2017ewj}.

As a closing remark, we would like to point out that, part of the interest in finding CP violation in the neutrino system, rests in the fact that CP-violating phases offer a simple and elegant way to explain the baryon asymmetry of the Universe via leptogenesis (although it would be possible to generate it other ways even if this phase, normally referred to as the Dirac phase, were zero). 
Singularly, CPT violation allows an even more simple and elegant way to do so, as in this case there is no need to fulfill the Sakharov conditions and the asymmetry can be generated in equilibrium, see for example Ref.~\cite{Barenboim:2001ac}.
In any case, the status of CPT as a fundamental symmetry of our theories should make it more subject to scrutiny and not act as a shield against it.

\section*{Acknowledgments}
GB acknowledges support from FPA2017-845438 and the Generalitat Valenciana under grant PRO\-ME\-TEOII/2017/033, also partial support from the European Union FP10 ITN ELUSIVES (H2020-MSCAITN-2015-674896) and INVISIBLES-PLUS (H2020- MSCARISE- 2015-690575).
CAT and MT are supported by FPA2017-85216-P(MINECO/AEI/FEDER, UE), PROMETEO/2018/165 (Generalitat Valenciana) and
FPA2017-90566-REDC (Red Consolider MultiDark).
CAT is supported by the FPI grant BES-2015-073593. 
MT acknowledges financial support from MINECO through the Ram\'{o}n y Cajal contract RYC-2013-12438.

\bibliographystyle{kp}
%\bibliography{bibliography}  

\begingroup\raggedright\endgroup

\end{document}